\documentclass[letter]{aa}

\usepackage{graphicx} 
\usepackage[varg]{txfonts}

\begin{document}

\title{Tracing the non-thermal pressure and hydrostatic bias \\ in galaxy clusters}
\titlerunning{$P_{NT}$ and $b$ in galaxy clusters}

\author{S. Ettori\inst{1,2} \and D. Eckert\inst{3} }
\institute{
        INAF, Osservatorio di Astrofisica e Scienza dello Spazio, via Piero Gobetti 93/3, 40129 Bologna, Italy
        \email{stefano.ettori@inaf.it}
        \and
        INFN, Sezione di Bologna, viale Berti Pichat 6/2, I-40127 Bologna, Italy
        \and
        Department of Astronomy, University of Geneva, Ch. d'Ecogia 16, CH-1290 Versoix, Switzerland
}
\authorrunning{Ettori \& Eckert}

\abstract{We present a modelization of the non-thermal pressure, $P_{NT}$, and we apply it to the X-ray (and Sunayev-Zel'dovich) derived radial profiles of the X-COP galaxy clusters.
We relate the amount of non-thermal pressure support to the hydrostatic bias, $b$, and speculate on how we can interpret this $P_{NT}$ in terms of the expected levels of
turbulent velocity and magnetic fields. 
Current upper limits on the turbulent velocity in the intracluster plasma are used to build a distribution $\mathcal{N}(<b) - b$, from which we infer that 50 per cent of local galaxy clusters 
should have $b < 0.2$ ($b<0.33$ in 80 per cent of the population).
The measured bias in the X-COP sample that includes relaxed massive nearby systems is 0.03 in 50\% of the objects and 0.17 in 80\% of them.
All these values are below the amount of bias required to reconcile the observed cluster number count in the cosmological framework set from {\it Planck}.
}


\maketitle

\section{Introduction}

Plasma in galaxy clusters and groups is an almost completely ionized gas, is a generator of magnetic fields, and is subjected to turbulent motions.
It is expected to thermalize after the accretion into the potential well on a timescale of the order of few gigayears, 
with a residual kinetic component whose distribution and total amplitude are still unknown. 

In this Letter, we propose a modelization of this residual non-thermal pressure, which supports the gas in equilibrium within an assumed dark-matter gravitation field, 
and apply the suggested technique to the observed thermodynamic profiles recovered for the objects in our {\it XMM-Newton} Cluster Outskirts Project (X-COP) sample \citep{xcop}.
We build the relations between this non-thermal pressure and any magnetic fields and/or turbulent velocity in the plasma and determine how the non-thermal pressure translates
into a hydrostatic bias on the mass measurement.

\section{The hydrostatic equation, the non-thermal pressure, and the hydrostatic bias}

Euler's equation for an ideal fluid (i.e. a fluid in which thermal conductivity and viscosity do not play a relevant role)
in a gravitational potential $\phi$ and with a velocity $\vec{v}$,  pressure $P_{\rm gas}$, and density $\rho_{\rm gas}$ 
is \citep[e.g.][]{landau59,suto13}
\begin{equation} 
\frac{\partial \vec{v}}{\partial t} +(\vec{v} \cdot \nabla)\vec{v} = - \frac{1}{\rho_{\rm gas}} \nabla P_{\rm gas} - \nabla \phi.
\label{eq:euler}
\end{equation}

\cite{lau13}, \cite{nelson14}, and \cite{biffi16} discuss the relative role of the different contributors to the total potential, allowing the total mass, $M_{\rm tot}$, to be written as the quantity that satisfies the following differential equation:
 \begin{equation} 
\frac{1}{\rho_{\rm gas}} \frac{dP_{\rm tot}}{dr} = -\frac{d\phi}{dr} = 
- \frac{G M_{\rm tot}}{r^2},
\label{eq:hee}
\end{equation}
where $G$ is the gravitational constant and $P_{\rm tot} = P_T +P_{NT}$ is the total pressure that we can express as the sum of the
thermal component ($P_T = P_{\rm gas} =\rho_{\rm gas} \, kT_{\rm gas}  / (\mu m_{\rm u}) = n_{\rm gas} \, kT_{\rm gas}$
\footnote{Throughout this Letter, $n$ is the electron density that relates to the gas density via the equation $\mu n_{\rm gas} = \mu_e n$, 
with $\mu \approx 0.6$ and $\mu_e \approx 1.16$})
and a non-thermal (not better defined) quantity, $P_{NT}$, that includes components generated from the velocity field, $\vec{v}$, in Eq.~\ref{eq:euler}
and accounts for: 
the rotational support due to mean tangential motions of gas; spatial variations in the mean radial streaming gas velocities; 
eventual correlations between the radial and tangential components of the random motions; 
and temporal variations in the mean radial gas velocities at a fixed radius.
These temporal variations, also known as the acceleration bias, have been shown in \cite{nelson14} to be less than few per cent in relaxed systems 
but comparable to the effect induced from turbulent and bulk motions in disturbed ones, with a contribution 
that spans values between 40 and 20 per cent of the non-thermal pressure support moving outwards \citep[see e.g.][]{biffi16,angelinelli20}.

If we write the total mass as $M_{\rm tot}  \sim r^2/n \, d P_{\rm tot}/dr$ and the mass component due to the thermal-only pressure
as $M_T \sim r^2/n \, d P_{T}/dr$, the following relation holds between the two
\begin{equation}
% M_{\rm tot} & \sim (r ^2/n) \, d P_{\rm tot}/dr  \sim M_T +(r^2/n) \, d P_{\rm NT}/dr  \nonumber \\
 % & \sim M_T +\alpha M_{\rm tot} +(r^2/n) P_{\rm tot} d\alpha/dr  \nonumber \\
% M_{\rm tot} (1-\alpha) & =  M_T \, (1 +P_{\rm tot} \frac{d\alpha/dr}{d P_T/dr})  \nonumber \\
M_{\rm tot} = M_T \frac{1 +P_{\rm tot} (d\alpha/dr)/(d P_T/dr)}{1 - \alpha} \, = \, M_T \frac{1 +A}{1 - \alpha},
\label{eq:mtot_mt}
\end{equation}
where 
\begin{eqnarray}
\alpha & = & \frac{P_{NT}}{P_{\rm tot}}, \nonumber \\
A & = & P_{\rm tot} \frac{d\alpha/dr}{d P_T/dr}.
\end{eqnarray}

We can now link this expression to the hydrostatic bias, $b$, the correction needed to reconcile hydrostatic mass measurements with the expected `true' value 
\citep[see e.g. the reviews in][for an extensive discussion on the measurements of $b$ and their cosmological implications]{ettori13,pratt19}:
\begin{equation}
M_{\Delta, T} =  (1-b) M_{\Delta, \rm tot},
\label{eq:mbias}
\end{equation}
where we consider the estimates of the mass at a given overdensity $\Delta = M_{\Delta} / (4/3 \pi \rho_{cz} R_{\Delta}^3$) with respect to
the critical mass density at redshift $z$, $\rho_{cz} = 3 H_z^2 / (8 \pi G)$, with 
$H_z = H_0 \left[ \Omega_{\rm m}  (1+z)^3 +1 -\Omega_{\rm m} \right]^{1/2}$ being the Hubble constant at that redshift 
in a flat Universe with $\Omega_{\rm m} = 1 - \Lambda$.
All the measurements quoted in the present analysis have been made assuming $H_0 = 70$ km s$^{-1}$ Mpc$^{-1}$ and $\Omega_{\rm m} = 0.3$.

In general, the computation of Eq.~\ref{eq:mbias} relies on the availability of  $M_{\Delta, T}$ from X-ray measurements and $M_{\Delta, \rm tot}$ from 
what is assumed to be a less biased true mass proxy, for example the one based on weak-lensing signals \citep[e.g.][]{meneghetti10}.
We note, however, that some assumptions have to be made on where (i.e. at which radius) the comparison is done to estimate $b$.
If the mass is estimated at a given overdensity, $\Delta$, or at the same physical radius, $r$, by combining Eqs.~\ref{eq:mtot_mt} and \ref{eq:mbias}, 
we can write\begin{equation}
b = b(r) = \frac{\alpha(r) +A(r)}{1+A(r)},
\label{eq:bias}
\end{equation}
which becomes $b = \alpha$ when $\alpha$ is a constant (i.e. $d\alpha/dr = 0$, and thus $A=0$).
Hereafter, we assume implicitly that $b$ depends on radius.
When needed, we will indicate to which radius (typically $R_{500}$) we are referring for some specific value of $b$.
In the case under consideration,  the mass values are estimated at the same physical radius (e.g. at the overdensity, $\Delta$, for the total mass profile, $M_{\rm tot}$, 
i.e. the radius $R_{\Delta, \rm tot}$), but only $M_{T}(R_{\Delta, T})$ is known, and $R_{\Delta, \rm tot}$ has to be inferred.
Then, by definition, the following relation holds between $M_{\Delta, \rm tot} = M_{\rm tot}(R_{\Delta, \rm tot})$ and $M_{\Delta, T} = M_{T}(R_{\Delta, T})$, 
$M_{\Delta, \rm tot} / R_{\Delta, \rm tot}^3 = M_{\Delta, T} / R_{\Delta, T}^3$, allowing us to write
\begin{eqnarray}
M_{\rm tot}(R_{\Delta, \rm tot}) & = & M_T(R_{\Delta, T}) \frac{R_{\Delta, tot}^3}{R_{\Delta, T}^3}  \nonumber \\
& = & M_T(R_{\Delta, T})  \, C_M(R_{\Delta, T}, R_{\Delta, tot}) \frac{1 +A}{1 - \alpha} \Big\rvert_{R_{\Delta, tot}}.
\label{eq:mdelta}
\end{eqnarray}

Under the assumption that we measure $M_T(R_{\Delta, T})$, the hydrostatic bias is estimated as
$(1-b) = R_{\Delta, T}^3 / R_{\Delta, tot}^3$ or $(1-b) = (1-\alpha)/(1+A) |_{R_{\Delta, tot}} / C_M(R_{\Delta, T}, R_{\Delta, tot})$,
where $C_M(R_{\Delta, T}, R_{\Delta, tot})$ is the ratio between the `thermal' mass evaluated at $R_{\Delta, tot}$ and at $R_{\Delta, T}$, 
which can be evaluated for a given mass profile of $f(r) = M_T(r) / M_{T0}$ as $C_M(R_{\Delta, T}, R_{\Delta, tot}) = f(R_{\Delta, tot}) / f(R_{\Delta, T})$.

\section{Modeling the non-thermal pressure}

In this section we investigate how these calculations can be done by adopting some functional forms for $P_T$ and $P_{NT}$.

As we have shown in \cite{ghirardini19poly}, a polytropic function, with an effective polytropic index $\gamma$, is a very good representation for the thermal gas distribution:
$P_T = P_{0T} \, (n/n_0)^{\gamma}$. 
Given our ignorance on the true, and probably not univocal, origin of the non-thermal pressure 
\citep[apart from speculations based on hydrodynamical simulations in e.g.][]{shi14,nelson14,angelinelli20}, 
we continue to adopt the thermal electronically charged gas distribution as a proxy for its distribution: $P_{NT} = P_{0NT} \, (n/n_0)^{\beta}$.
Then, 
\begin{equation}
\alpha = \frac{P_{NT}}{P_{\rm tot}} =  \left[ \frac{P_{0T}}{P_{0NT}} \left(\frac{n}{n_0}\right)^{\gamma-\beta}+1\right]^{-1}
\label{eq:alpha}
\end{equation}
\begin{equation}
A = P_{\rm tot} \frac{d\alpha/dr}{d P_T/dr} =  \left( \frac{\beta}{\gamma} -1 \right) \alpha,
\label{eq:aa}
\end{equation}
where we have made use of the following relations: $dP_{T} / dr = P_{0T} \gamma n^{\gamma-1} dn/dr$ and  $dP_{NT} / dr = P_{0NT} \beta n^{\beta-1} dn/dr$. 

As final step, we can now also rewrite the hydrostatic bias in Eq.~\ref{eq:bias} as
\begin{equation}
b = \frac{\alpha +A}{1+A} = \frac{\alpha \beta}{(1-\alpha) \gamma +\alpha \beta} = \frac{\beta P_{NT}}{\gamma P_T + \beta P_{NT}}.
\label{eq:b_a}
\end{equation}
Again, when $d\alpha/dr = 0$, then $\gamma = \beta$, and $b = P_{NT} / (P_T +P_{NT}) = \alpha$.
For physical reasons, $\alpha \ge 0$ and $\gamma>0$. Then, $b \ge 0$ only when $\beta \ge 0$.
If $d\alpha/dr \ne 0$, then we have a complete modelization of the radial behaviour of the hydrostatic bias, $b$.

To constrain the non-thermal component, \cite{shi16} propose adopting an analytic model \citep[described in][]{shi14} that is able to predict 
the amplitude of non-thermal pressure and its radial, mass, and redshift dependences, but only once the mass accretion history of the cluster is known.
Alternatively, and under the common condition that thermodynamic profiles based on only X-rays and Sunayev-Zel'dovich \citep[SZ;][]{sz} effect are available, 
some external information on the total gravitational potential is needed.
For instance, the total mass profiles can be derived from gravitational lensing data \citep[see e.g.][]{sayers21} or from stellar kinematics \citep[as in early-type galaxies; see e.g.][]{churazov08,humphrey13}.
\cite{ff13} fix the gas fraction at the viral radius equal to the cosmic baryon fraction, with the contribution from stars subtracted, 
to model any extra contribution in pressure given X-ray derived profiles.
To account for the complex physical processes that can affect the relative distributions of the baryons in a cosmological contest, 
we \citep[][]{ghirardini18,eckert19} propose using a `universal' gas fraction, $f_{g, U}$, derived from a mean baryon fraction measured 
at some given overdensities (typically $\Delta=$500 and 200) in massive haloes extracted from hydrodynamical simulations and hence corrected 
for a statistical contribution of the stellar fraction (i.e. $f_{g, U} = f_{\rm b, sims} - f_{\rm star, stat}$).
In the following analysis, we adopt the measurements of this universal gas fraction presented and discussed in our previous work
\citep[see e.g. Sect. 3.1 in][]{eckert19}: $f_{g, U} (R_{500}) = 0.131 \pm 0.009$ and $f_{g, U} (R_{200}) = 0.134 \pm 0.007$.
We discuss below the impact of this assumption on our constraints of the non-thermal pressure.

In \cite{eckert19}, a model for $\alpha$ (instead of $P_{NT}$) -- presented in \cite{nelson14} as a simple, not physically motivated, universal functional form of the 
radial behaviour observed in cosmological hydrodynamical simulations -- is adopted and solved iteratively in its two free parameters (a third scale parameter is maintained fixed). 
In the present work, we suggest adopting a modelization of the non-thermal pressure as directly proportional to the electron density -- which is more physically plausible since the electron density is the natural reservoir of the magnetic fields and particles to be accelerated -- and
with model parameters that are more easily interpretable as physical quantities.

The total true mass is then recovered by comparing $f_{g, U}$ to the observed $f_{g, T}$, which is based on the `thermal-only' gravitational mass profile \citep[described in][]{ettori19} 
and on the gas mass profile previously corrected for some gas clumping that might bias the gas density reconstruction 
(see the discussion on the recovered thermodynamic profiles in \citealt{ghirardini19} and Sect.~5.1 in \citealt{eckert19} on the systematic uncertainties affecting $f_{g, T}$).
By propagating the hydrostatic bias in Eqs.~\ref{eq:mbias} and \ref{eq:bias}, we can write
\begin{equation}
f_g = \frac{M_g}{M_{\rm tot}} \Big\rvert_{R_{\Delta, tot}} = \frac{M_{g, T}}{M_T} \Big\rvert_{R_{\Delta, tot}} \, \frac{1 -\alpha}{1 +A} \Big\rvert_{R_{\Delta, tot}} .
\end{equation}
We note that the gas fraction will be evaluated at two different radii, $R_{\Delta, tot}$ and $R_{\Delta, T}$, which correspond to the true and thermal-only mass profile, respectively.
With only $R_{\Delta, T}$  known from direct observations ($R_{\Delta, tot}$  depends on the mass profile corrected for the hydrostatic bias), 
we compute all these quantities at $R_{\Delta, T}$ and write
\begin{equation}
\frac{M_g}{M_{\rm tot}} \Big\rvert_{R_{\Delta, T}} = f_g \Big\rvert_{R_{\Delta, tot}} \, C_f(R_{\Delta, tot}, R_{\Delta, T}) = 
  f_{g, T}\Big\rvert_{R_{\Delta, T}} \, \frac{1-\alpha}{1 - \alpha (\gamma -\beta)/\gamma }\Big\rvert_{R_{\Delta, T}}.
\label{eq:fgas}
\end{equation}
In this equation: $f_g \rvert_{R_{\Delta, tot}}$ is extracted from hydrodynamical simulations; $f_{g, T}\rvert_{R_{\Delta, T}}$ and $\gamma$ are measured from the observations; and
$\alpha$ and $\beta$ are the unknowns.

The correction factor $C_f(R_{\Delta, tot}, R_{\Delta, T}) = f_g(R_{\Delta, T}) / f_g(R_{\Delta, tot})$ used to convert the gas fraction from numerical simulations at different radii 
can be estimated by assuming that $n \sim r^{-\epsilon}$ in the region of interest (around $R_{500}$ and $R_{200}$).
Then, $M_g(R_{\Delta, T}) / M_g(R_{\Delta, tot}) = (R_{\Delta, T}/R_{\Delta, tot})^{3-\epsilon}$, 
and, using Eq.~\ref{eq:mdelta}, $C_f = f_g(R_{\Delta, T})/f_g(R_{\Delta, tot}) \approx (R_{\Delta, T}/R_{\Delta, tot})^{3-\epsilon} \times C_M(R_{\Delta, T}, R_{\Delta, tot})$.
Moreover, by assuming the power-law behaviour of $n$ and $P$, and applying the HE to $M_T$ and $M_{\rm tot}$ as in Eq.~\ref{eq:mtot_mt},
we obtain that
\begin{eqnarray}
M_T = r^{1 +\epsilon -\epsilon \gamma}  & \nonumber \\
M_{tot} = r^{1 +\epsilon -\epsilon \gamma} & \left( 1+\frac{\beta}{\gamma} \frac{P_{0NT}}{P_{0T}} r^{\gamma \epsilon -\beta\epsilon} \right),
\label{eq:mass_plaw}
\end{eqnarray}
which can be used to estimate $C_M(R_{\Delta, T}, R_{\Delta, tot}) = M_{tot}(R_{\Delta, tot}) / M_{tot}(R_{\Delta, T})$.
Using the relation between $R_{\Delta, tot}$ and $R_{\Delta, T}$ (see Eq.~\ref{eq:mdelta}) with Eq.~\ref{eq:mass_plaw}, we can then write
\begin{equation}
R_{\Delta, T}^{\epsilon-\gamma\epsilon -2} = R_{\Delta, tot}^{\epsilon-\gamma\epsilon -2} \left(
 1+\frac{\beta}{\gamma} \frac{P_{0NT}}{P_{0T}} R_{\Delta, tot}^{\gamma \epsilon -\beta\epsilon} \right)
\end{equation}
and, finally,
\begin{equation}
C_f = \frac{f_g(R_{\Delta, T})}{f_g(R_{\Delta, tot})} = \left( \frac{R_{\Delta, T}}{R_{\Delta, tot}} \right)^{-\epsilon} \left(
 1+\frac{\beta}{\gamma} \frac{P_{0NT}}{P_{0T}} R_{\Delta, T}^{\gamma \epsilon -\beta\epsilon} \right)^{-1}.
\label{eq:c_f}
\end{equation}

In the present analysis, because we do not use the profiles in numerical simulations, we assume $C_f = 1$. 

Considering that we know (i) the gas fraction, both the true and universal one,  $f_{g, U}$, and the observed thermal one, $f_{g, T}$, 
at (at least) two given points ($R_{500}$ and $R_{200}$, from the thermal mass profile, $M_T$),
(ii) the gas density, $n$, and  (iii) a (mean) polytropic index, $\gamma$, in that radial range, we can estimate $\alpha$ and $\beta$, 
measuring both the normalization $P_{0NT}$ and the radial dependence of the modelled $P_{NT}$. 
In details, if we define $F_1 = f_{g, U}/f_{g, T}|_{r_1}$, $F_2 = f_{g, U}/f_{g, T}|_{r_2}$, and $\gamma = \hat{\gamma}$ as estimated from 
the recovered density and pressure profiles, and if we assume that the unknowns 
$p_0 = P_{0T}/P_{0NT}$ and $\beta$ do not vary with radius (though this assumption can also be relaxed once more measurements of $F_i$ are available), 
we can write $\alpha_i = P_{NT} / P_{\rm tot}|_{r_i} =  \left[ p_0 \, (n_i/n_0)^{\hat{\gamma}-\beta}+1\right]^{-1}$ and the following system of two equations with the two unknowns:
\begin{equation}
    \begin{cases}
      F_1 = \frac{1-\alpha_1}{1+\alpha_1 (\beta-\hat{\gamma})/\hat{\gamma}} \\
      F_2 = \frac{1-\alpha_2}{1+\alpha_2 (\beta-\hat{\gamma})/\hat{\gamma}}
      \end{cases}\,.
\label{eq:f1f2}
\end{equation}

\begin{figure}
\includegraphics[width=0.5\textwidth, keepaspectratio,clip,trim=2cm 1.5cm 1cm 8cm]{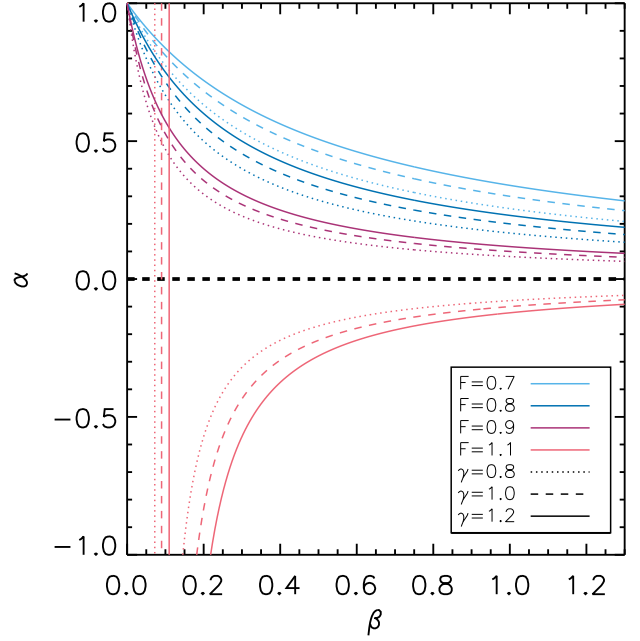}
\caption{Representation of Eq.~\ref{eq:f1f2} in the form $\alpha = (1-F) / \left[ 1+F (\beta-\gamma)/\gamma \right]$
for different values of $F = f_{g, U}/f_{g, T}$ and of the effective polytropic index, $\gamma$.
} \label{fig:ai-bb}
\end{figure}

We note that the impact of the assumption on the value of $f_{g, U}$ at a given radius can be easily quantified from Eq.~\ref{eq:f1f2}.
We represent this function by plotting in Fig.~\ref{fig:ai-bb} how $\alpha$ varies as a function of $\beta$ for different measurements of the effective polytropic index, $\gamma$, 
and of the ratios between the expected and the observed gas mass fraction, $F$. 
As expected, $\alpha$ increases as $F$ decreases, for example by assuming lower values of the universal gas fraction, $f_{g, U}$, for a given measured $f_{g, T}$.
Estimates of $\alpha$ between 0 and 1 are obtainable for any $\beta$ only when $F<1$, with $\alpha \rightarrow 0$ when $F \rightarrow 1$.
For values of $F>1$, $\alpha$ will be larger than zero only when $\beta <  \gamma (F-1)/F$.

\begin{table*}[ht]
\caption{Parameters of the non-thermal pressure model for the X-COP clusters.} 
\setlength\tabcolsep{2.2pt}
\begin{center} \begin{tabular}{ccccccc} \hline
Cluster & $\alpha_{500}/\alpha_{200}$ (E19) & $F_{500} / F_{200}$ & $\gamma$ & $P_{0NT}/P_{0T}$ & $\beta$ & $\alpha_{500}/\alpha_{200}$ (this work) \\ \hline
A1644 & $10.5_{-10.5}^{+0.0} / 14.8_{-14.8}^{+0.0}$ & $1.023 \pm 0.095 / 1.063 \pm 0.108$ & $1.117 \pm 0.066$ &  - & -  & $0.0^{+0.0}_{+0.0} / 0.0^{+0.0}_{+0.0}$ \\
A1795 & $2.2_{-2.2}^{+5.6} / 6.7_{-4.5}^{+6.0}$ & $0.942 \pm 0.073 / 0.931 \pm 0.058$ & $1.090 \pm 0.096$ & 1.79e-02 & 0.937 & $6.6^{+3.8}_{-2.0} / 8.0^{+12.6}_{-5.6}$ \\
A2029 & $6.0_{-5.7}^{+5.8} / 10.4_{-10.4}^{+9.0}$ & $0.929 \pm 0.072 / 0.882 \pm 0.058$ & $1.249 \pm 0.068$ & 1.51e-03 & 0.749 & $11.3^{+5.0}_{-3.5} / 18.3^{+11.0}_{-10.1}$ \\
A2142 & $15.8_{-4.8}^{+4.5} / 18.6_{-8.8}^{+7.1}$ & $0.829 \pm 0.063 / 0.798 \pm 0.050$ & $1.252 \pm 0.060$ & 3.80e-02 & 1.038 & $19.9^{+2.9}_{-2.3} / 23.4^{+7.6}_{-5.6}$ \\
A2255 & $5.6_{-5.6}^{+6.8} / 6.1_{-6.1}^{+6.3}$ & $0.856 \pm 0.085 / 0.918 \pm 0.123$ & $1.173 \pm 0.040$ & 4.68e+00 & 1.597 & $11.0^{+8.0}_{-2.3} / 6.2^{+17.1}_{-3.3}$ \\
A2319 & $43.6_{-3.6}^{+3.5} / 52.3_{-4.6}^{+3.4}$ & $0.693 \pm 0.056 / 0.565 \pm 0.041$ & $1.304 \pm 0.043$ & 1.20e-03 & 0.484 & $54.4^{+11.8}_{-7.3} / 67.4^{+12.5}_{-10.7}$ \\
A3158 & $8.5_{-5.8}^{+5.7} / 12.5_{-11.6}^{+8.9}$ & $0.903 \pm 0.076 / 0.865 \pm 0.072$ & $1.194 \pm 0.055$ & 6.59e-03 & 0.838 & $13.2^{+5.8}_{-6.0} / 18.3^{+14.0}_{-13.3}$ \\
A3266 & $11.2_{-11.2}^{+0.0} / 15.9_{-15.9}^{+0.0}$ & $0.992 \pm 0.096 / 1.241 \pm 0.217$ & $0.899 \pm 0.031$ &  - & -  & $0.0^{+0.0}_{+0.0} / 0.0^{+0.0}_{+0.0}$ \\
A644 & $3.2_{-3.2}^{+6.4} / 5.6_{-5.6}^{+6.4}$ & $0.992 \pm 0.113 / 0.964 \pm 0.116$ & $1.356 \pm 0.118$ &  - & -  & $0.0^{+14.6}_{+3.9} / 0.0^{+26.6}_{+1.1}$ \\
A85 & $10.2_{-5.6}^{+4.9} / 11.5_{-9.5}^{+8.9}$ & $0.873 \pm 0.067 / 0.843 \pm 0.058$ & $0.871 \pm 0.045$ & 5.20e-02 & 0.735 & $14.7^{+3.4}_{-2.7} / 18.1^{+8.7}_{-6.3}$ \\
RXC1825 & $5.1_{-5.1}^{+5.1} / 15.2_{-7.8}^{+6.4}$ & $0.985 \pm 0.077 / 0.865 \pm 0.060$ & $0.984 \pm 0.050$ &  - & -  & $0.0^{+28.9}_{+9.8} / 0.0^{+57.8}_{+14.1}$ \\
ZW1215 & $11.9_{-11.9}^{+0.0} / 15.7_{-15.7}^{+0.0}$ & $1.236 \pm 0.126 / 1.457 \pm 0.162$ & $0.999 \pm 0.053$ &  - & -  & $0.0^{+0.0}_{+0.0} / 0.0^{+0.0}_{+0.0}$ \\
\hline \end{tabular} \end{center}
\tablefoot{For each object, we quote: the constraints on $\alpha$ at $R_{500}$ and $R_{200}$ both in \cite{eckert19} (second column) and in the present work (last column);
the estimated $F_i$; the effective polytropic index, $\gamma$, measured in the radial range $[0.3, 1.6] R_{500}$; the parameters $1/p_0 = P_{0NT}/P_{0T}$ and
$\beta$, from which $\alpha$ is estimated (see Eq.~\ref{eq:alpha}).\ The quoted errors are at $1 \sigma$ c.l., and the quoted errors on $\alpha$ are obtained from 100 Monte Carlo runs 
by propagating the statistical errors on the observed gas density and pressure profiles, and the observed gas fractions.
} \label{tab:gamma_beta}
\end{table*}

\begin{figure}
\includegraphics[width=0.5\textwidth, keepaspectratio,clip,trim=1cm 1cm 1cm 7cm]{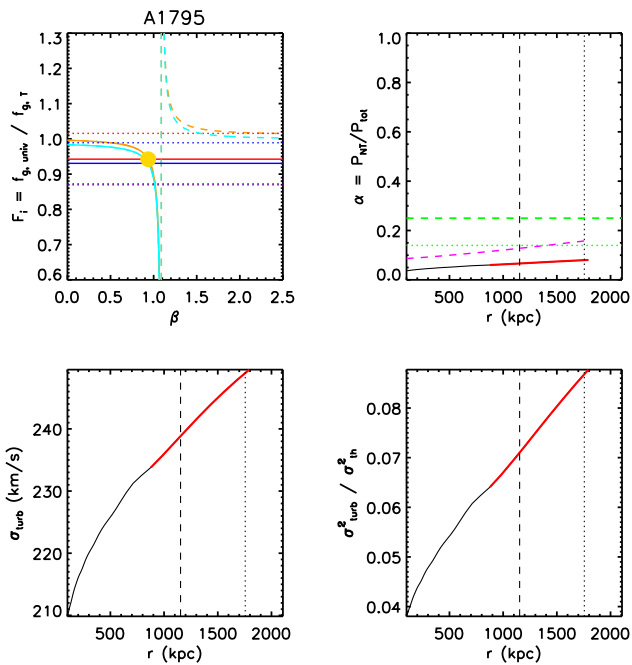}
\caption{
Example of the analysis applied to A1795. Upper left: Distribution of the ratio of the expected gas fraction (the dashed orange line  represents $R_{500}$ and dashed cyan $R_{200}$) and 
observed gas fraction, $F_i$ (solid horizontal lines, with errors represented by dotted lines; red represents $R_{500}$ and blue $R_{200}$) as a function of the slope, $\beta$ (see Eq.~\ref{eq:f1f2}); 
the best-fit value is indicated by the yellow dot. Upper right: Distribution of $\alpha$ as a function of radius with, overplotted in purple, the prediction from the model of $\alpha_{\rm turb}$ in \cite{angelinelli20} and the expectations
on the level of turbulence assuming $\eta=1/3$ (green dashed line) and magnetic energy $C_E=0.05$ (green dotted line; see Sect.~4 for details). Lower left: Distribution of the turbulent velocity as a function of radius (see Eq.~\ref{eq:sigma}).\ Lower right:  Distribution of the relative energy (Eq.~\ref{eq:ratioE}) as a function of radius.
In red, we show the region used to constrain $\gamma$; $R_{500}$ and $R_{200}$ are indicated with vertical dashed and dotted lines.
} \label{fig:best-fit}
\end{figure}

By dividing $F_1$ by $F_2$, we can solve for $p_0$, included in $\alpha_i$, as a function of $\beta$ and after some math obtain
\begin{equation}
p_0 = \frac{\beta}{\hat{\gamma}} \left[ \left( \frac{n_1}{n_2} \right)^{\hat{\gamma}-\beta} -\frac{F_1}{F_2} \right] \left(\frac{n_1}{n_0}\right)^{\beta-\hat{\gamma}} 
\left( \frac{F_1}{F_2} -1 \right)^{-1}.
\label{eq:p0}
\end{equation}
Then, we solve numerically using an array of values for $\beta$ and by inserting Eq.~\ref{eq:p0} into the second term of one of the two equations in Eq.~\ref{eq:f1f2}, 
as represented in Fig.~\ref{fig:best-fit}.
We present the results for the X-COP objects in Table~\ref{tab:gamma_beta}.
All our estimates are consistent with the results presented in \cite{eckert19}.
For five clusters, the constraints on $\alpha$ are consistent with zero: three of them have an estimated $F_i > 1$ at $R_{500}$ and/or $R_{200}$, 
implying a negative hydrostatic bias. 

We note that our algorithm allows the radial behaviour of $\alpha$ to be constrained without any assumptions on the underlying functional form 
(such as those suggested from recent numerical simulations; see e.g. \citealt{shaw10,nelson14,angelinelli20}) 
and on a single object basis, although a universal mean gas fraction has to be imposed on all the systems.

\section{Discussion on the interpretation of $P_{NT}$}

The properties of the intracluster plasma are regulated through the action of cosmological supersonic flows,
which accrete mass onto the cluster halo and inject energy that is then dissipated over different scales and under different forms, 
including non-thermal plasma components such as relativistic particles and magnetic fields \citep[see e.g.][]{bj14}.
\cite{miniati15} present evidence from numerical simulations that the energy components follow a hierarchy that 
can be represented through some general relations between thermal energy, $E_T$, energy in turbulence, $E_{\rm turb}$, and energy in magnetic field $E_B = B^2 / (8\pi)$:
\begin{equation}
\frac{E_T}{E_{\rm turb}} = 3 \left( \frac{\eta}{1/3} \right)^{-1}, \hspace*{0.5cm}
\frac{E_T}{E_B} = 60 \, (\gamma -1) \, \left( \frac{\eta}{1/3} \right)^{-1} \left( \frac{C_E}{0.05} \right)^{-1},
\end{equation}
where $\eta$ is the fraction of thermal energy originating from turbulent dissipation and $C_E$ is the fraction of turbulent (kinetic) energy that converts into magnetic energy; 
$\eta$ and $C_E$ have expected values of about 1/3 and 0.05, respectively.

From the above equations, we can write that the magnetic field, $B$, is related to the thermal pressure, $P_T$, as 
\begin{equation}
\frac{B^2}{8\pi} = \frac{\eta  \,  C_E}{\gamma-1} \, P_T.
\end{equation}
Using the polytropic law for $P_T = P_{0T} (n/n_0)^{\gamma}$, we can then write $B = (8 \pi \,  \eta \,  C_E \, P_{0T})^{0.5} (\gamma-1)^{-0.5} (n/n_0)^{\gamma/2} = B_0 (n/n_0)^{\gamma_B}$.
With this modelization, we estimate a median $B_0$ of about 15 $\mu$G at 100 kpc.
Furthermore, we can write the magnetic pressure as a fraction of the total pressure as
\begin{equation}
\frac{P_B}{P_{tot}} = \frac{\eta C_E / (\gamma-1) \, P_T}{P_T +P_{NT}} =   \frac{\eta C_E / (\gamma-1) \, P_T}{P_T + \eta P_T}
= \frac{\eta C_E}{(\gamma-1) \, (1+\eta)}.
\label{eq:pb}
\end{equation}
In the extreme case that the entire non-thermal pressure support is in the form of turbulence, then $P_{NT} = E_{\rm turb}$ and
$\eta = P_{NT}/P_T = \alpha / (1-\alpha)$, or $\alpha = \eta / (\eta+1)$ , which we overplot as a dashed green line in Fig.~\ref{fig:best-fit} for $\eta=1/3$.
We indicate the fraction of magnetic energy for $C_E =0.05$ with a dotted green line.
From our dataset of measurements of $\alpha(R_{500})$, by requiring that $P_B \equiv P_{NT}$, we obtain median (mean) values of 
$\eta$ of about 0.15 (0.3) and $C_E$, from the inversion of Eq.~\ref{eq:pb}, $\sim$ 0.19 (0.16), with the former being more than half of, and
the latter three to four times larger than, the estimates from numerical simulations.

Finally, if we convert $P_{NT}$ into the velocity dispersion associated with the turbulence of the intracluster medium (ICM), we can write
\begin{equation}
P_{NT} = a \, \rho \, \sigma_{\rm turb}^2 = a \, \mu_e m_p n \, \sigma_{\rm turb}^2
,\end{equation}
which implies
\begin{equation}
\sigma_{\rm turb}^2 = \frac{P_{0NT} \, (n/n_0)^{\beta-1}}{a \, \mu_e m_p} = \frac{\alpha \, P_T}{1-\alpha} \frac{1}{a \, \mu_e m_p n} 
\label{eq:sigma}
,\end{equation}
where $a=$1/3 and 1 for the isotropic value and the radial component only, respectively, and $\rho$ is the gas mass density equal to $\mu_e m_p n$ 
\citep[e.g.][]{vazza18,angelinelli20}.
Similar treatment for the determination of the non-thermal pressure support is discussed in \cite{humphrey13}, where the constraints from 
stellar dynamics are used to provide an unbiased estimate of the true gravitational potential in elliptical galaxies.

With the thermal velocity dispersion defined as $\sigma_{\rm th}^2 =  k_B T / (a \mu_e m_p) = P_T /(a \mu_e m_p n)$, we can write the ratio between 
the turbulent velocity dispersion and the velocity dispersion produced by thermal motions as
\begin{equation}
\frac{\sigma_{\rm turb}^2}{\sigma_{\rm th}^2} = \frac{\alpha}{1-\alpha}.
\label{eq:ratioE}
\end{equation}
Inverting this equation, we can write the hydrostatic bias as a function of the turbulent velocity,
\begin{equation}
b = \left( \frac{\sigma_{\rm th}^2}{\sigma_{\rm turb}^2} +1 \right)^{-1}.
\label{eq:b_ratioE}
\end{equation}

\begin{figure}
  \includegraphics[width=0.5\textwidth,keepaspectratio,clip,trim=2.5cm 1.5cm 0 9cm]{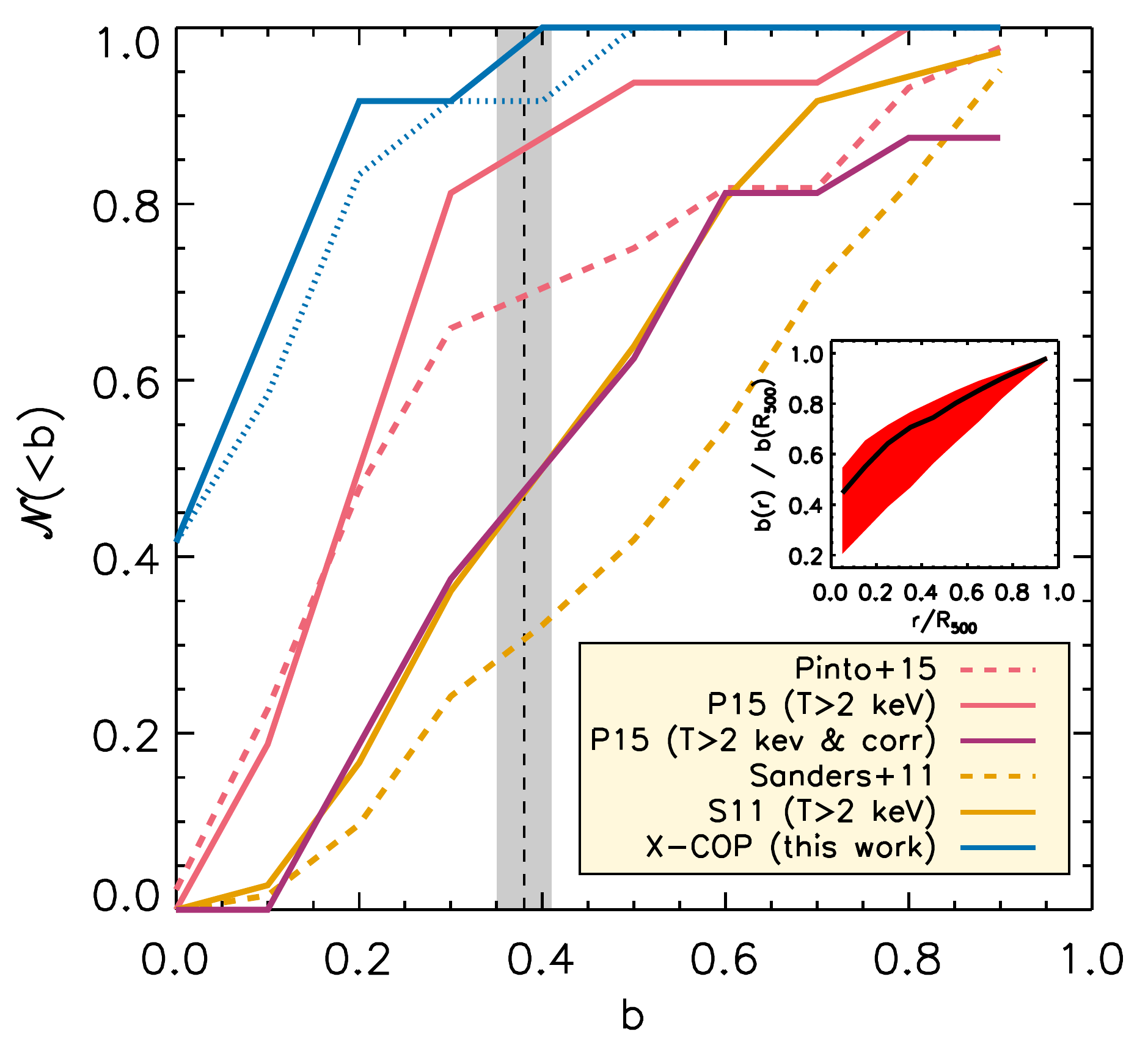}
\caption{
Density distribution of the hydrostatic bias, $b$.
The values of $b$ are inferred from Eq.~\ref{eq:b_a}, using the values tabulated in Table~\ref{tab:gamma_beta},
for the X-COP objects at $R_{500}$ (solid line) and at $R_{200}$ (dotted line), and from Eq.~\ref{eq:b_ratioE} for the upper limits for the clusters 
in \cite{pinto15} (16 out of 44 with $T> 2$ keV; upper limits estimated within 0.8 arcmin) and \cite{sanders11} (36 out of 62).
The purple line shows the constraints that include the correction to extrapolate the estimates from the core to $R_{500}$  
(see text and the inset, where we plot median distribution, with first and third quartiles, of the hydrostatic bias in the X-COP sample as a function of radius).
The shaded region shows the constraints obtained from the {\it Planck} collaboration by combining the cosmic microwave background anisotropies and
the cluster number count \citep{planck_cosmo}. 
} \label{fig:vturb_bias}
\end{figure}

We used published values of upper limits on $\sigma_{\rm turb}$ to reconstruct the expected upper limits on $b$ (from Eq.~\ref{eq:b_ratioE}, $b$ decreases as $\sigma_{\rm turb}$ decreases).
In Fig.~\ref{fig:vturb_bias}, we show the $\mathcal{N}(<b) - b$ (or hydrostatic bias density) distribution based on the constraints presented in \cite{sanders11} and \cite{pinto15} and obtained
from the width of emission lines in {\it XMM-Newton} Reflection Grating Spectrometer spectra extracted from the inner regions of a sample of 62 galaxy clusters, groups, and elliptical galaxies and 
the 44 nearby, bright objects in the CHEERS sample, respectively.
We note that the upper limits on $b$ are larger in cooler systems. Limiting our analysis to the clusters with $kT>2$ keV, and interpolating
the distribution measured in the hotter systems of the CHEERS sample, we estimate upper limits on $b < (0.33, 0.19, 0.09)$ in 80, 50, and 20 per cent of the population. 
Considering that these measurements are limited to the cluster's cores, we estimated a median correction on the bias $b$, estimated at different radii, in the X-COP sample 
(see inset in Fig.~\ref{fig:vturb_bias}) and obtain $b(0.1 R_{500}) \approx 0.5 b(R_{500})$, with a general increase in the bias moving outwards.
Applying this correction to the upper limits in \cite{pinto15}, we evaluate $b < (0.64, 0.41, 0.20)$ in 80, 50, and 20 per cent of the population, respectively.

We can now compare these upper limits with the results we obtain for the X-COP sample (see Fig.~\ref{fig:vturb_bias}).
The distribution of the bias in our objects covers lower values of $b$ (with 80\% of the objects with $b<0.17$ and 50\% with $b<0.03$ at $R_{500}$), 
confirming that our bright local systems are, on average, relaxed (i.e. closer to the assumption of a spherically symmetric gas distribution in hydrostatic equilibrium with the
underlying gravitational potential), as they were selected at origin.
Three objects studied in our work are also present in  \cite{pinto15}.
From the 1 $\sigma$ upper limits quoted there, we infer that $b = \alpha$ from turbulence of about $<$11\%, $<$51\%, and $<$29\% for A85, A1795, and A2029, respectively, 
Only for A85 does this value appear to be similar to the constraints presented in Table~\ref{tab:gamma_beta}; 
for A2029 and A1795, the levels of bias induced from the current upper limits are more than two and seven times the constraints we obtain.

Overall, these distributions of the hydrostatic bias, $b$, at $R_{500}$ are in good agreement with the mean values of about 10-20 per cent obtained in
hydrodynamical simulations of massive haloes in a cosmological context \citep[see e.g.][]{biffi16,pearce20}.

\section{Conclusions}

We have presented a new model of the non-thermal pressure as $P_{NT} = P_{0NT} \, (n/n_0)^{\beta}$ 
and have applied it to the X-COP results on the radial profiles of the gas density, temperature, and pressure 
to constrain the parameters $P_{0NT}$ and $\beta$.
We converted this non-thermal pressure support to the bias $b$ that affects the reconstruction of the mass profile through the hydrostatic equilibrium equation.
We also translated $P_{NT}$ into expected contributions in the form of magnetic fields and turbulent velocity in the ICM.
Regarding turbulent velocity, we put interesting constraints on the expected distribution of the hydrostatic bias using upper limits on the turbulent velocities available in the literature.
We propose the hydrostatic bias density distribution $\mathcal{N}(<b)-b$ as a way to assess statistically interesting and useful quantities that have an impact 
on the use of galaxy clusters for astrophysical and cosmological purposes.
We show that current constraints (both upper limits on the turbulent velocity in the ICM and measurements from the gas mass fraction in X-COP; see Fig.~\ref{fig:vturb_bias})
suggest that most of the studied galaxy clusters have a $b$ lower than (or marginally consistent with, when the upper limits extrapolated up to $R_{500}$ are considered) 
the value of $b=0.38 (\pm 0.03)$ required to reconcile the measurements of the cosmic microwave background anisotropies with 
the SZ-based cluster number count from  {\it Planck} data \citep{planck_cosmo}.
We note that to accommodate for this value of $b$ with the median estimates of $\beta$ (0.84) and $\gamma$ (1.19) recovered in the X-COP sample, 
by using Eqs.~\ref{eq:b_a} and \ref{eq:f1f2} and the measured $f_{g,T}$, we require $f_{g,U}(R_{500})$ to be 62\% of the adopted value of 0.13 (i.e. 
a universal gas fraction of about 0.08). This value is more typical for $10^{14} M_{\odot}$ systems, where the action of feedback 
has a larger impact than in more massive, X-COP-like, objects on the expected distribution of baryons, at least as it is resolved in state-of-the-art 
cosmological hydrodynamical simulations \citep[see e.g. Fig.~15 in][]{eckert21}.

Future X-ray instruments with improved spectral resolution (e.g. XRISM\footnote{\url{https://xrism.isas.jaxa.jp/en/}} and X-IFU\footnote{\url{http://x-ifu.irap.omp.eu/}} 
on board Athena\footnote{\url{https://www.the-athena-x-ray-observatory.eu/}}) will tighten these limits, providing significant constraints on the 
turbulent velocity and allowing the picture of the energy and mass budget of galaxy clusters to be completed.
In the meantime, projects such as {\it CHEX-MATE}\footnote{\url{http://xmm-heritage.oas.inaf.it/}} will provide
a direct measurement of $b$ by combining different mass proxies in an SZ-selected representative
sample of galaxy clusters.

\begin{acknowledgements}
We thank the referee for some valuable comments.
SE acknowledges financial contribution from the contracts ASI-INAF Athena 2019-27-HH.0,
``Attivit\`a di Studio per la comunit\`a scientifica di Astrofisica delle Alte Energie e Fisica Astroparticellare''
(Accordo Attuativo ASI-INAF n. 2017-14-H.0), INAF mainstream project 1.05.01.86.10, and
from the European Union's Horizon 2020 Programme under the AHEAD2020 project (grant agreement n. 871158).
\end{acknowledgements}

\bibliographystyle{aa}
\bibliography{p_nt_b}

\end{document}